\documentstyle[12pt]{article}

\catcode `\@=11
\@addtoreset{equation}{section}
\def\theequation{\arabic{section}.\arabic{equation}}
\catcode `\@=12

\newcommand{\be}{\begin{equation}}
\newcommand{\en}{\end{equation}}
\newcommand{\bea}{\begin{eqnarray}}
\newcommand{\ena}{\end{eqnarray}}
\newcommand{\beano}{\begin{eqnarray*}}
\newcommand{\enano}{\end{eqnarray*}}
\newcommand{\bee}{\begin{enumerate}}
\newcommand{\ene}{\end{enumerate}}

\newcommand{\D}{{\cal D}}
\newcommand{\s}{{\cal S}}
\newcommand{\Lag}{{\cal L}}
\newcommand{\R}{R \!\!\!\! R}
\newcommand{\N}{N \!\!\!\!\! N}

\begin{document}

\thispagestyle{empty}
 
\vspace*{1cm}

\begin{center}
{\Large \bf Multiplications of Distributions in one dimension and a First Application to
Quantum Field Theory}   \vspace{2cm}\\

{\large F. Bagarello}
\vspace{3mm}\\
  Dipartimento di Matematica ed Applicazioni, 
Fac.Ingegneria, Universit\`a di Palermo, I - 90128  Palermo, Italy\\
E-mail: Bagarell@unipa.it
\vspace{2mm}\\
\end{center}

\vspace*{1cm}

\begin{abstract}
\noindent 
	In a previous paper we have introduced a
class of multiplications of distributions in one dimension. Here we furnish different
generalizations of the original definition and we discuss some applications of these procedures to
the multiplication of delta functions and to quantum field theory.

\vspace{0.5cm}
\noindent
{\bf R\'esum\'e} Nous avons introduit dans une publication pr\'ec\'edente une classe de
multiplications entre distributions $\grave{a}$ une dimension. 
Nous donnons ici des g\'en\'eralisations diff\'erentes des d\'efinitions originelles et nous
discutons des applications de ces m\'ethodes $\grave{a}$ la multiplication des functions delta et
$\grave{a}$ la th\'eorie quantique des champs.

 \end{abstract}

\vfill

\newpage

\section{Introduction}

In the past years many attempts have been done to extend the ordinary
multiplication between functions to distributions. Of course, as for the
extensions of (almost) any kind, the procedure is not unique and, in fact, many
inequivalent proposals are nowadays present in the literature, see
\cite{fish,breme,zhi} among the others.

In this paper we generalize a class of multiplication of distributions introduced in a previous
work by the author, see \cite{bag1}. The original definition was based on two 
different regularizations of distributions, the analytic and the sequential
completion methods. In particular, this last procedure makes reference to
functions in $\D(\R)$ which generate the so-called delta-families. A delta
family is essentially a set of functions which approximate $\delta(x)$ in the
topology of $\D(\R)$. Sometimes, whenever the applications require it, it may
appear necessary to use a weaker form of this procedure.  Possible weakening of
the requirements in \cite{bag1} are part of the containt of this paper. In
particular, in Section 2 we relax some of the requirements given in
\cite{bag1}, so that, in principle, more distributions can be multiplied
between themselves, while in  Section 3 we give two
inequivalent multiplications among more than two distributions.

 The necessity for extending the multiplication to more than two distributions 
follows directly from physical examples: this is what we need to do whenever we
try to regularize three or four-points Green's functions in a given quantum
field model. In fact, it has been recognized since Wightmann's work, \cite{wig},
that the field operators are not operator-valued functions, but rather
distributions defined on a certain domain dense in a Hilbert space. We know also
that a field theory is often defined via a lagrangian density, $\Lag$, which
depends on the products of such fields considered at coincident points. This is,
of course, an operation which has no rigorous mathematical meaning in this
naive form. One of the most famous consequences of this procedure is that certain
Feynman diagrams diverge, and that analogous divergences are observed also in
many matrix elements of the dynamical variables. Many attempts have been made
in the past decades to give a rigorous meaning to quantum field theory (QFT).
This has generated essentially two different approaches: constructive QFT, as
proposed first by Wightmann, where the Wightmann functions, that is some 'matrix
elements' of the fields, are the relevant dynamical variables, and the
algebraic QFT, developped by Haag and Kastler in the sixties, which we will not
consider here. In \cite{wig} it is widely discussed a sort of regularization
procedure for the Wightmann functions: they can be recovered as the boundary
values of some holomorphic functions. Nevertheless, even using this
regularization, the problem of the divergences in a perturbative QFT still
exists and the solution proposed are, in our opinion, unsatisfactory.

A possible way out is to compute these Feynman graphs using the regularizations of the fields,
instead of the fields themselves, and to remove the regularization only at the end. This is the
path we will follow in Section 5. The results we will obtain, however, show
once more the difficulty of the problem: in particular we will see that our
regularization procedure, as it is, is not powerful enough to avoid the
appearance of the divergences in a free QFT in 1+1 dimensions.

An idea close to our is also behind Colombeau's book, \cite{col}, where a
systematic approach to quantum field theory (QFT) is proposed. The lack of
uniqueness in the regularization procedure makes Colombeau's work not
resolutive. For instance, other approaches, more along the lines of this paper,
can be found in \cite{breme,breme2}.

The paper is divided as follows:

in the next Section we start recalling the definition of the multiplication
given in \cite{bag1}. We take also the opportunity for briefly discussing some
new result. Then we 'relax' this definition of the multiplication to better deal
with physical models involving distributions of $\s'$, like in QFT;

in Section 3 we discuss two possible generalizations of the theory discussed in Section 2 to more
than two distributions;

in Section 4 we show many examples involving delta functions of both the
generalizations proposed. We also discuss some physical examples;

in Section 5 we apply our procedure  to an easy quantum
field model, the free $1+1$ Klein-Gordon theory;

in the last Section, finally, we comment the results and state a plane for the future.

\section{Definition of the multiplication}

In this Section we briefly recall, for readers' convenience, the basic 
definitions and results of the multiplication introduced in \cite{bag1}. We
will slightly modify this  definition in a second time, so to build up a
 framework  which is more suitable for applications to QFT.

Let ${\cal V}$ be the subspace of all the functions in $C^{\infty}$ with
arbitrary support, ${\cal E}$, with the following properties:

$i) \hspace{1.1cm} \phi (x)\; |x| \leq k_0 \mbox{   for} \; |x| \rightarrow
\infty, $

$
ii) \hspace{1cm}  \phi^{(n)}(x)\; |x| \leq k_n \mbox{  for} \; x \rightarrow
\infty, $

where $k_0, k_1,...$ are constants. The convergence is defined as in ${\cal
E}$.

Let ${\cal V'}$ be the dual space of ${\cal V}$. For distributions in this
space it has been shown in \cite{breme} that the function
\be
 {\bf T}^0(z) \equiv \frac{1}{2\pi i}\, T\cdot(x-z)^{-1} 
\label{21}
\en
exists and is holomorphic in $z$ in the whole $z$-plane minus the support of $T$. The
function 
\be
T_{red}(x,\epsilon) \equiv {\bf T}^0(x+i\epsilon)-{\bf T}^0(x-i\epsilon)
\label{22}
\en
is further weakly convergent to the distribution $T$ when $\epsilon$ goes to zero,
 \cite{breme}. Also, if $T(x)$ is a continuous function with compact support, then 
$T_{red}(x,\epsilon)$ converges uniformly to $T(x)$ on the whole real axis for 
$\epsilon \rightarrow 0^+$.

The other ingredient of the multiplication in \cite{bag1} is the 
method of the sequential completion, which makes reference to the so-called
$\delta$-sequences. In \cite{bag1} we have called  $\delta$-sequence a sequence
of functions $\delta_n(x)\equiv n\phi(nx)$, where $\phi \in {\cal D}({\bf R})$
is a given function with supp $\phi \subseteq [-1,1]$ and $\int \phi (x) \, dx
=1$.  Then, $\forall \,  T \in {\cal D'}({\bf R})$, the  convolution $T_n \equiv
T*\delta_n$ is a $C^{\infty}-$function, for any fixed $n\in {\bf N}$. The
sequence $T_n$ converges to $T$ in the topology of ${\cal D'}$, when $n \rightarrow
\infty$. Moreover, if $T(x)$ is a continuous function with compact support
then $T_n(x)$ converges uniformly to $T(x)$.

In \cite{bag1} we have proceeded in the following way:

\noindent
for any couple of distributions $T,S \, \in
{\cal V'}, \, \forall \, \alpha, \beta >0$ and $\forall \, \Psi \, \in {\cal
D}$ we have defined the following quantity: 
\be 
(S\otimes T)_n^{(\alpha,\beta)}(\Psi) \equiv \frac{1}{2} \int_{-\infty}^{\infty}
[S_n^{(\beta)}(x)\,  T_{red}(x,\frac{1}{n^\alpha}) + T_n^{(\beta)}(x)\,
S_{red}(x,\frac{1}{n^\alpha})]\, \Psi (x) \, dx  
\label{ddd}
\en 
where $S_n^{(\beta)}(x) \equiv (S*\delta_n^{(\beta)})(x),$ with $\delta_n^{(\beta)}(x)
\equiv n^{\beta} \Phi (n^{\beta}x)$.

\noindent
Hence, the two distributions $S$ and $T$ in ${\cal V'}$ are said to be multipliable
if the limit of $(S\otimes T)_n^{(\alpha,\beta)}$ for $n\rightarrow \infty$ exists
finite in a weak sense. Finally, we have defined

\be
(S\otimes T)_{(\alpha,\beta)} (\Psi )\equiv \lim_{n \rightarrow \infty}
(S\otimes T)_n^{(\alpha,\beta)}(\Psi ).
\label{def}
\en
 In \cite{bag1} we have proved, among other things, that this product extend the usual
product of the functions, in the sense that if $T(x)$ and $S(x)$ are two continuous
functions with compact supports then the product $T_n^{(\beta)}(x)\,
S_{red}(x,\frac{1}{n^\alpha})$  converges uniformly to $T(x)\, S(x)$.   
As mathematical applications of our definition we have discussed the
possibility of multiplying two (derivatives of) delta functions localized at the same
point. Here we want to make this information complete. In particular
we want to extend the multiplication to arbitrary derivatives of $\delta(x)$, $\delta^{(k)}$.
Since this result is a straightforward generalization of what has been done in
\cite{bag1}, we will not give all the details. If we want to define $(\delta^{(k)}
\otimes \delta^{(l)})_{(\alpha,\beta)}$ then we are forced to consider different situations
depending on the parity of the integers $k$ and $l$. We get     
\bea
    (\delta^{(k)} \otimes \delta^{(l)})_{(\alpha,\beta)}  = \left\{
        \begin{array}{ll}
        0  & \alpha>(k+l+2)\beta \\
        \frac{(k+l+1)!}{\pi}A_{k+l+2} \delta, &  \alpha=(k+l+2)\beta  \:\:k,l\mbox{ even}\\
        -\frac{(k+l+1)!}{\pi}A_{k+l+2} \delta, &  \alpha=(k+l+2)\beta  \:\:k,l\mbox{ odd}\\
        0   &       \alpha=(k+l+2)\beta, \: \: k \mbox{ even and } l
        \mbox{ odd}, 
       \end{array}
        \right. 
\label{23}
\ena

\noindent
where we have taken, as in \cite{bag1}, 
\bea
    \Phi(x)  = \left\{
          \begin{array}{ll}
            \frac{x^m}{F} \cdot \exp\{\frac{1}{x^2-1}\},    &       |x| <1 \\
           0,   &       |x| \geq 1. 
       \end{array}
        \right. 
\label{23bis}
\ena
Here $F$ is a normalization constant, and we have defined, whenever they exist,
$A_j \equiv \int_{-\infty}^{\infty} \frac{\Phi(x)}{x^j}\, dx$, for integers $j$. In
order to have a finite regularization (\ref{23}), we always have to choose a function $\Phi(x)$
with $m$ even and such that $m>k+l+1$. It is interesting to notice that equation
(\ref{23}) implies, among the others, the following equalities: 
\beano
  \left\{
   \begin{array}{ll}
  &(\delta'' \otimes \delta'')_{(\alpha,\beta)} = -(\delta' \otimes
\delta''')_{(\alpha,\beta)} = (\delta \otimes \delta'''')_{(\alpha,\beta)} \\
   &(\delta' \otimes \delta')_{(\alpha,\beta)}= - (\delta \otimes
\delta'')_{(\alpha,\beta)}, \\
       \end{array}
        \right. 
\enano
 which show that, at least for this particular example, the usual property
of the derivatives of the distributions are satisfied by the
product $ \otimes _{(\alpha,\beta)}$. 

\vspace{3mm}

As we have already discussed in the Introduction, we are interested in applying our
proposal of regularization and multiplication to quantum fields, which are operators
whose matrix elements belong to $\s'$. It is therefore natural to generalize a bit the
above definition, trying to construct a framework more directly
related to the physics. In particular, we
 modify the definition of the sequential completion, which is strongly
related, in its original version, to distributions in $\D'$. There exist also technical
reasons which suggest to relax the definition of the sequential completion. We will
comment on this point in Section 5. 

Using the terminology of \cite{kor}, we call delta
sequence of Dirichelet type a family of functions $\delta_k(x)$, which satisfy the
following conditions: \beano &i)& \int_{-A}^{A} \delta_k(x) \, dx \rightarrow 1 \mbox{
when } k \rightarrow \infty \mbox{  for a certain } A>0;\\
& ii)& \forall \gamma>0, \: \forall f\in {\cal L}^1(\R) \mbox{ then } 
\lim_{k \rightarrow \infty} (\int_{-\infty}^{-\gamma} + 
\int_{\gamma}^{\infty})(\delta_k(x)\, f(x)\, dx) = 0; \\ 
& iii)& \exists \, C_1, C_2, \mbox{ positive constants }: \:  |\delta_k(t)| \leq
\frac{C_1}{|t|}+C_2. 
\enano

In particular, it is an easy exercise to prove that for any $\Phi(t)$ belonging
to $\s$, such that $\int_{-\infty}^\infty\Phi(t) \, dt =1$, then the family of functions
$\delta_n^{(\beta)}(x) \equiv n^\beta \Phi (n^\beta x)$ generates a delta
sequence of Dirichelet type if $\beta>0$.

We have the following:
\vspace{4mm}

\noindent
	{\bf Proposition 1}.

 Let $T \in \s'({\bf R})$ and $\delta_n^{(\beta)}(x)$ be a delta sequence of Dirichelet 
functions. Then the convolution $T_n^{(\beta)} \equiv T*\delta_n^{(\beta)}$ is a
$C^{\infty}-$function, for any fixed $n\in {\bf N}$. The sequences
$\delta_n^{(\beta)}(x)$ and  $T_n^{(\beta)} (x)$ converge respectively to $\delta$ and
to $T$ in the topology of $\s'$, when $n \rightarrow \infty$, for all $\beta >0$.

Moreover, if $T(x)$ is a H$\ddot o$lder-continuous function with compact support 
$[a,b]$, then $T_n^{(\beta)}(x)$ converges uniformly to $T(x)$ on every interior subinterval of
$[a,b]$.  \vspace{4mm}

\noindent
	{\underline {Proof}}

The $C^{\infty}-$ nature of the function $T_n^{(\beta)}(x)$ is a well known property of the
convolutions, which can be found, for instance, in \cite{kir}. 

In reference \cite{kor} it is proved that any $\delta-$sequence of Dirichelet type
weakly converges to the $\delta$ function with respect to any function which has
a finite derivative in the origin. A fortiori therefore $\delta_n^{(\beta)}
$ will converge to $\delta$  in the topology of $\s'$. From this fact it easily
follows the convergence of $T_n^{(\beta)}$ to $T$ in $\s'$.

The last statement is again contained in \cite{kor}.
\hfill $\Box$

  \vspace{3mm}

\noindent{\bf Remarks }.--
(a) Another generalization of the $\delta$-family is also discussed in \cite{kor}. The
family is now called {\em delta family of positive type} and, as the name itself
suggests, its functions must all be not negative. This is, in general, a
 strong requirement which is not necessarily satisfied by the 'generating' function
$\Phi(t)$ we will use in the application to QFT, and this is the reason why we have
focused our attention to Dirichelet's type functions. Nevertheless, even for such a delta family a
Proposition like the one above can be stated; minor differences are required in the
hypotheses but the results, essentially, coincide.

(b) Any $\Phi(t) \in \D(\R)$ generating a 'standard' delta family also generates a
delta family of Dirichelet type and, if $\Phi(x) \geq 0$ for all $x\in \R$, also a
delta family of positive type. 

(c) One may wonder why we have introduced so many families of delta functions: the
reason is that the choice of the function $\Phi$ cannot be made in general a priori by us, but it
is often forced by the model which has to be regularized. In particular, in the example of
QFT we will show that there is no reason, in general, for $\Phi(x)$ to have compact
support or to be positive.

(d)  Proposition 1 can be used to show that the new multiplication still extends the
usual multiplication of continuous functions with compact support, in the sense that
if $T(x)$ and $S(x)$ are H$\ddot o$lder-continuous functions with compact support in
$[a,b]$ then, $\forall \alpha, \beta>0$ and $\forall \, \Psi \in \s$, then 
$$ (T\otimes S)_{(\alpha,\beta)}(\Psi) =
\int_{-\infty}^{\infty} T(x)\, S(x) \, \Psi (x)\, dx.
$$

\vspace{5mm}

The definition of the multiplication is now, formally, the same as in
eqs. (\ref{ddd}) and (\ref{def}). The only difference is in the mathematical nature of
$\delta_n^{(\beta)}$. 

 We end this Section giving the extension of
the definition of multiplication to the case in which the
distributions $S$ and $T$ do not commute, even if we will not meet with this
problem in this paper. In this condition we are forced to symmetrize
the original definition (\ref{ddd}), (\ref{def}). Let $S$ and $T$ be two operator
valued distributions. Keeping the same notation as before, we define  \bea 
(S\otimes T)_{(\alpha,\beta)}(\Psi) \equiv \frac{1}{4}
\lim_{n\rightarrow \infty} \int_{-\infty}^{\infty} [S_n^{(\beta)}(x)\, 
T_{red}(x,\frac{1}{n^\alpha}) \!\!\!&+& \!\!\! T_n^{(\beta)}(x)\,
S_{red}(x,\frac{1}{n^\alpha})+ \nonumber \\ 
+T_{red}(x,\frac{1}{n^\alpha})\, S_n^{(\beta)}(x) + S_{red}(x,\frac{1}{n^\alpha})\,
T_n^{(\beta)}(x) \!\!\!&]&\!\!\! \!\!\, \Psi (x) \, dx.  
\label{24}
\ena
Of course this definition must be understood in the weak (Hilbert) sense. Moreover,
whenever $S$ and $T$ commute (again in the weak (Hilbert) sense), the above definition
returns the original one.

\section{Multiplying More Distributions}

Up to now we have focused our interest to the multiplication of two distributions and its
possible definitions. This is not enough in many physical situations, like, for
instance, in the computation of the four-points Green's functions in a scalar $\lambda
\varphi^4$ theory, \cite{itz}. In this perspective we will now analyze possible
extensions of the definition (\ref{def}) when more than two
distributions are considered. In particular we will suggest two different, inequivalent,
approaches, and we will discuss some examples. Which method  has to be chosen only depends
on which one gives theoretical results in (a better) agreement with the experiments (or
with the common sense). We will return on this point with an example at the end of the next
Section.

In this paper we will consider only commuting distributions. This is an useful condition to
simplify all formulas.

The first method we are going to discuss is, in our opinion, the most natural one since
it does not need any new ingredient for its definition. We start with two
distributions $S_1$ and $S_2$. Their multiplication, if it exists, is defined by
(\ref{def}). Let us now suppose to be interested in defining the product of three
distributions  $S_1$, $S_2$ and $S_3$ in ${\cal V'}$. It is quite natural to consider the following
quantity
\bea
\left( S_1 \otimes S_2 \otimes S_3\right)_n^{(\alpha,\beta)} &\equiv & \nonumber \\
\frac{1}{3} {\Large [}
(S_1\otimes S_2)_n^{(\alpha,\beta)}\,S_3 \!\!\!&+ & \!\!\!(S_1\otimes
S_3)_n^{(\alpha,\beta)}\,S_2+ (S_2\otimes S_3)_n^{(\alpha,\beta)}\,S_1 {\Large ]},
\label{31}
\ena
which is certainly well defined for any fixed $n$, since any term above is the
product of a $C^\infty$ function for a distribution. As usual, what may or may not exist
is the limit for $n\rightarrow \infty$ of $\left( S_1 \otimes S_2 \otimes
S_3\right)_n^{(\alpha,\beta)}(\Psi)$, for any $\Psi \in \D(\R)$. If this limit exists we say that
the distributions can be multiplied and we put
\be
\left( S_1 \otimes S_2 \otimes S_3\right)_{(\alpha,\beta)}(\Psi) \equiv \lim_{n\rightarrow
\infty} \left( S_1 \otimes S_2 \otimes S_3\right)_n^{(\alpha,\beta)}(\Psi).
\label{32}
\en

It is useful to notice that formula (\ref{31}) would look rather more complicated without
the working hypothesis of the commutativity of the distributions. 

Let us now try to define a
multiplication between four distribution. In this case, of course, we cannot repeat the
same steps leading to equation (\ref{32}), since the quantity $\left( S_1 \otimes S_2
\otimes S_3\right)_{(\alpha,\beta)}S_4$  would necessarily contain the product of two
un-regularized distributions. We have to define this multiplication in a different way.
This problem can be easily overcome simply by coupling the distributions in all the
possible ways and then using twice the regularization. This implies that the
product of four distributions should depend on four indices, two $\alpha$'s and two
$\beta$'s. Explicitly we have: 
\bea
\hspace{-6mm}& &\left(  S_1 \otimes S_2 \otimes S_3 \otimes S_4
\right)_{(\alpha_1,\alpha_2,\beta_1,\beta_2)}(\Psi) \equiv \lim_{n\rightarrow
\infty} \frac{1}{6} 
 {\Large \{ }(S_1\otimes S_2)_n^{(\alpha_1,\beta_1)}(S_3\otimes
S_4)_n^{(\alpha_2,\beta_2)} +\nonumber \\
\hspace{-6mm}& & +(S_1\otimes S_3)_n^{(\alpha_1,\beta_1)}(S_2\otimes
S_4)_n^{(\alpha_2,\beta_2)}+ (S_1\otimes S_4)_n^{(\alpha_1,\beta_1)}(S_2\otimes
S_3)_n^{(\alpha_2,\beta_2)} + \nonumber \\
\hspace{-6mm}& & + (\alpha_1,\beta_1) \leftrightarrow (\alpha_2,\beta_2){\Large \}}(\Psi),  
\label{32bis}
\ena
whenever this limit exists. To be more explicit, for instance the first term of this formula reads
\bea
&&\left[(S_1\otimes S_2)_n^{(\alpha_1,\beta_1)}(S_3\otimes S_4)_n^{(\alpha_2,\beta_2)}\right]
(\Psi) = \nonumber \\
&&=\int_{-\infty}^\infty S_{1,n}^{(\beta_1)}(x)\, 
S_{2,red}(x,\frac{1}{n^{\alpha_1}}) S_{3,n}^{(\beta_2)}(x)\,  S_{4,red}(x,\frac{1}{n^{\alpha_2}})
\, \Psi(x) \,dx \label{32tris}
\ena

The multiplication of five distributions is now naturally defined
in analogy with the one in (\ref{31}) and (\ref{32}). We put 
\beano
& &\left(  S_1 \otimes S_2 \otimes S_3 \otimes S_4  \otimes S_5
\right)_{(\alpha_1,\alpha_2,\beta_1,\beta_2)}(\Psi) \equiv \nonumber \\
& & \frac{1}{5} {\Large [} (S_1 \otimes S_2 \otimes S_3 \otimes
S_4)_{(\alpha_1,\alpha_2,\beta_1,\beta_2)}S_5 + (S_1 \otimes S_2 \otimes S_3 \otimes
S_5)_{(\alpha_1,\alpha_2,\beta_1,\beta_2)}S_4 + \nonumber \\
& &(S_1 \otimes S_2 \otimes S_4 \otimes
S_5)_{(\alpha_1,\alpha_2,\beta_1,\beta_2)}S_3 + (S_1 \otimes S_3 \otimes S_4 \otimes
S_5)_{(\alpha_1,\alpha_2,\beta_1,\beta_2)}S_2 + \nonumber \\
& &(S_2 \otimes S_3 \otimes S_4 \otimes S_5)_{(\alpha_1,\alpha_2,\beta_1,\beta_2)}S_1
{\Large ]}(\Psi),
\enano
whenever the right hand side exists for any $\Psi(x) \in \D(\R)$.

It is clear now how this procedure can be generalized to the multiplication of an
arbitrary number $N$ of distributions:

\noindent
whenever $N$ is even we have to proceed like in (\ref{32bis}), that is we consider
all the different $N/2$ pairs of distributions, regularize each pair, and then try
to remove the regularization. If $N$ is odd, we simply have to multiply one un-regularized
distribution with the regularization of the even $N-1$ remaining ones.

In all the examples discussed in this work we will stick to the situation in which all the
distributions coincide. In this case all the formulas are strongly simplified. Whenever
the limits below exist we have:
\bea
\hspace{-8mm}( S \otimes S )_{(\alpha,\beta)} \!\!\!&= &\!\!\! \lim_{n\rightarrow \infty} 
S_n^{(\beta)}(x)\,  S_{red}(x,\frac{1}{n^\alpha}) 
\label{33} \\
\hspace{-8mm}( S \otimes S \otimes S )_{(\alpha,\beta)} \!\!\!&= &\!\!\!  S \, \lim_{n\rightarrow
\infty}  ( S \otimes S )_n^{(\alpha,\beta)}
\label{34} \\
\hspace{-8mm}( S \otimes S \otimes S \otimes S)_{(\alpha_1,\alpha_2,\beta_1,\beta_2)} \!\!\!&=&
\!\!\! \lim_{n\rightarrow \infty} ( S \otimes S )_n^{(\alpha_1,\beta_1)} ( S \otimes S
)_n^{(\alpha_2,\beta_2)}
\label{35}\\
\hspace{-8mm}( S \otimes S \otimes S \otimes  S \otimes S)_{(\alpha_1,\alpha_2,\beta_1,\beta_2)}
\!\!\! &=&\!\!\!  S \, \lim_{n\rightarrow \infty} ( S \otimes S )_n^{(\alpha_1,\beta_1)} ( S
\otimes S )_n^{(\alpha_2,\beta_2)},
\label{36}
\ena
and so on. The generalization to a bigger number of distributions is straightforward. All
the formulas above are obviously thought in their week forms: they must be {\em applied} to a
generic function $\Psi \in \D$, like in equation (\ref{32tris}).

\vspace{4mm}

We now discuss a different proposal which again extends the multiplication 
introduced in (\ref{def}) for two distributions.

First of all, let us introduce two complex quantities $a_1, a_2$, with $a_1+a_2=2$. We
modify the original definition (\ref{def}) of the multiplication of two distributions by
saying that {\em two distributions $S_1$ and $S_2$ are A-multipliable if there exists a
choice of $a_1$ and $a_2$, with $a_1+a_2=2$, such that the following limit exists}: 
\bea
\hspace{-14mm}&&( S_1 \otimes S_2 )^A_{(\alpha,\beta)}(\Psi) \equiv \nonumber \\
\hspace{-14mm}&&\equiv\!\!\frac{1}{2} \lim_{n\rightarrow
\infty}\int_{-\infty}^\infty\left[ a_1 S_{1,n}^{(\beta)}(x)\,  S_{2,red}(x,\frac{1}{n^\alpha}) +a_2
S_{1,red}(x,\frac{1}{n^\alpha}) \, S_{2,n}^{(\beta)}(x) \right]\!\Psi(x)\,dx,
\label{37}
\ena
for any $\Psi \in \D(\R)$.

Of course definition (\ref{def}) turns out to be simply a special case of this one when we
take $a_1=a_2=1$. It is interesting to notice that our new multiplication depends now not only on
$\alpha, \beta,$ but also on $a_1$ and $a_2$. Of course, it may happen that one
contribution in (\ref{37}) does not converge for $n\rightarrow \infty$. In this case,
while the multiplication in (\ref{def}) is not defined, the one above still exists for a
clever choice of $a_1$ and $a_2$.

The length of the formulas rapidly increases when the number of distributions to
be multiplied grows up. Already for three distributions we need to introduce six
parameters, $a_{11}, a_{12},a_{13},a_{21},a_{22}$ and $a_{23}$, whose sum must be equal to
$6$. The A-multiplication of the three distributions is said to exist if there exists a
choice of the coefficients $a_{ij}$'s and of the pair $(\alpha, \beta)$ such that the limit
below exists: 
\bea 
\!\!\!\!\!\!\!& &( S_1 \otimes S_2 \otimes S_3 )^A_{(\alpha,\beta)}(\Psi) \equiv \frac{1}{6} 
\lim_{n \rightarrow \infty} \int_{-\infty}^\infty {\Large [} a_{11} S_{1,n}^{(\beta)}(x)
S_{2,n}^{(\beta)}(x) S_{3,red}(x,\frac{1}{n^\alpha}) + \nonumber \\
\!\!\!\!\!\!\!\!\!\!\! & & + a_{12}
S_{1,n}^{(\beta)}(x) S_{2,red}(x,\frac{1}{n^\alpha}) S_{3,n}^{(\beta)}(x) +  a_{13}
S_{1,red}(x,\frac{1}{n^\alpha}) S_{2,n}^{(\beta)}(x) S_{3,n}^{(\beta)}(x)+\nonumber \\
\!\!\!\!\!\!\!\!\!\!\!& &+ a_{21} S_{1,n}^{(\beta)}(x) S_{2,red}(x,\frac{1}{n^\alpha})
S_{3,red}(x,\frac{1}{n^\alpha}) + a_{22}  S_{1,red}(x,\frac{1}{n^\alpha}) 
S_{2,n}^{(\beta)}(x) S_{3,red}(x,\frac{1}{n^\alpha})+\nonumber \\
\!\!\!\!\!\!\!\!\!\!\!& & + a_{23}  S_{1,red}(x,\frac{1}{n^\alpha})
S_{2,red}(x,\frac{1}{n^\alpha}) S_{3,n}^{(\beta)}(x){\Large ]}\, \Psi(x)\, dx, 
\label{38}
\ena
for all $\Psi \in \D$.

In the case of four distributions the number of the coefficients grows up to $14$, so that
it is more and more difficult to correctly keep into account all these contributions.
However, the situation drastically simplifies when all the distributions coincide. In
this case we have symmetry reasons which give some extra conditions on the coefficients
$a$. 

For instance, in the case of two equal distributions, from definition (\ref{37}) it is
evident that we have to take $a_1=a_2=1$. Hence, this method returns the usual result, see
equation (\ref{33}). 

From (\ref{38}) we deduce that, if $S_1=S_2=S_3=S$, then necessarily
$a_{11}=a_{12}=a_{13}=:b_1$ and $a_{21}=a_{22}=a_{23}=:b_2$, and therefore $b_1+b_2=2$.
Consequently (\ref{38}) becomes now 
\bea
\hspace{-18mm}&&( S \otimes S \otimes S )^A_{(\alpha,\beta)}(\Psi)= \nonumber \\
\hspace{-18mm}&&=\!\frac{1}{2} \lim_{n \rightarrow
\infty}  \int_{-\infty}^\infty\left[ b_1 (S_{n}^{(\beta)}(x))^2 S_{red}(x,\!\frac{1}{n^\alpha})\! 
+\! b_2 S_{n}^{(\beta)}(x) (S_{red}(x,\!\frac{1}{n^\alpha}))^2 \right]\Psi(x)\, dx.
\label{39}
\ena
Finally, without going into details, it is possible to prove that for the multiplication
of four equal distributions we need to introduce three parameters $c_1, c_2$ and $c_3$, such that
$2c_1+3c_2+2c_3=7$. The multiplication turns out to be
\bea
\hspace{-13mm}& &(S \otimes S \otimes S \otimes S)^A_{(\alpha,\beta)}(\Psi)=  \frac{1}{7}
\lim_{n \rightarrow \infty}\int_{-\infty}^\infty {\Large [} 2c_1 (S_{n}^{(\beta)}(x))^3
S_{red}(x,\frac{1}{n^\alpha}) + \nonumber \\ 
\hspace{-13mm}& &+ 3c_2 (S_{n}^{(\beta)}(x))^2
(S_{red}(x,\frac{1}{n^\alpha}))^2 + 2c_3 S_{n}^{(\beta)}(x) (S_{red}(x,\frac{1}{n^\alpha}))^3 
{\Large ]}\,\Psi(x)\,dx. \label{310}
\ena
The same procedure can be repeated even for a bigger number of distributions but we will
omit this generalization here since the difficulty grows up very fast  with the number
of distributions.

Just a comment before ending this Section: in our opinion, this last method appears to be less
natural than the first one. Nevertheless, we will
show in Section 4 that it works well in some examples, and its extra
degrees of freedom may, in turn, be useful in future applications. The main difference 
within the two methods proposed in this Section is that in the first one we increase the
number of indices $\alpha$ and $\beta$, while in the second one we keep this number
unchanged but we introduce new extra parameters which were not originally present in the
definition we gave in \cite{bag1}. As we have already observed, the preference must be
given to that method whose results are closer to the experimental data, or the the common
wisdom.

\section{Examples: delta functions}

We devote this Section to show how the multiplications defined previously work explicitly. In
particular, we will show that both methods proposed allow to define the product of an arbitrary
number of delta function in one dimension localized at the same point. The technique we are going
to use is very much the same as the one used in \cite{bag1} where two (derivatives of) delta
functions have been shown to be multipliable. In particular we will need very often the
well known Lebesgue dominated convergence theorem (LDCT), see \cite{reed} for example.

Before starting with the computation of the multiplications we remind the readers the
expressions of the two regularizations of the delta function, \cite{bag1}. We have:
\be
\delta_n^{(\beta)}(x) \equiv n^{\beta} \Phi (n^{\beta}x), \quad \beta>0
\label{41}
\en
and
\be
\delta_{red}(x,\frac{1}{n^\alpha})  = \frac{1}{\pi n^\alpha}\frac{1}
{(x^2+\frac{1}{n^{2\alpha}})} \quad \alpha>0.
\label{42}
\en

We begin with considering the first method proposed, eqs. (\ref{33})-(\ref{36}), taking
$S=\delta$. We fix first the form of the function generating the delta sequence. In this Section we
will always assume that $\Phi(x)$ is the one given in (\ref{23bis}), where $F$ is a given
normalization constant (of course $m$-depending) and $m$ is an integer which must be taken even
so to prevent $\int_{-1}^1\Phi(x)\, dx$ to be zero.

The result for $( \delta \otimes \delta )_{(\alpha,\beta)}$ is already contained in Section 2,
see (\ref{23}). Changing a little bit the notation for future convenience, we have, for any $\Psi
\in \D(\R)$, \be
    (\delta \otimes \delta)_{(\alpha,\beta)}(\Psi)  = \left\{
          \begin{array}{ll}
          \frac{1}{\pi}A_{1,2} \delta(\Psi),    &       \alpha=2\beta \\
           0,   &       \alpha>2\beta, 
       \end{array}
        \right. 
\label{43}
\en
where we have defined
\be
A_{i,j}\equiv \int_{-1}^{1} \frac{\left(\Phi(t)\right)^i}{t^j} \, dt.
\label{44}
\en
Of course, due to the presence of $A_{1,2}$ in $( \delta \otimes \delta
)_{(\alpha,\beta)}(\Psi)$, we need to take $m\geq 2$. Otherwise the integral defining $A_{1,2}$
would be divergent. 

This result coincides for both the methods proposed: this is obvious since the
different multiplications introduced in the last Section \underline{both} generalize the
multiplication discussed in \cite{bag1} and refined in Section 2.

It is very easy to compute the product of three delta functions using our recipe; equation
(\ref{34}) becomes now
\beano
(\delta \otimes \delta \otimes \delta)_{(\alpha,\beta)}(\Psi) \!\!\!\!&=& \!\!\!\!
\lim_{n\rightarrow \infty} \int_{-\infty}^{\infty} \delta(x) \, \delta_n^{(\beta)}(x)\,
\delta_{red}(x,\frac{1}{n^\alpha})\, \Psi (x) \, dx = \\
& &=\frac{1}{\pi} \Phi(0) \lim_{n\rightarrow \infty} n^{\alpha+\beta} \Psi(0)=0
\enano
since $\Phi(0)=0$ for any $m>0$, for any choice of $\alpha$ and $\beta$ in $\R_+$. 

Let us now move to the multiplication of four delta functions. The situation is no longer so
easy. Using (\ref{35}) we have \beano
& &(\delta \otimes \delta \otimes \delta \otimes \delta)_{(\alpha_1,\alpha_2,\beta_1,\beta_2)}
(\Psi) =  \\
& &= \lim_{n\rightarrow \infty} \int_{-\infty}^{\infty} \delta_n^{(\beta_1)}(x)\,
\delta_{red}(x,\frac{1}{n^{\alpha_1}}) \delta_n^{(\beta_2)}(x)\,
\delta_{red}(x,\frac{1}{n^{\alpha_2}}) \, \Psi (x) \, dx.
\enano
Here we are interested to show that there exists a choice of $m$, $\alpha_i$ and $\beta_i$
for which the limit of the right hand side of this equation exists finite. We will show
that such a result can be obtained already if we take $\alpha_1=\alpha_2=:\alpha$ and
$\beta_1=\beta_2=:\beta$, with some extra conditions on $\alpha$ and $\beta$. We call 
$(\delta \otimes \delta \otimes \delta \otimes \delta)_{(\alpha,\beta)}(\Psi) \equiv (\delta
\otimes \delta \otimes \delta \otimes \delta)_{(\alpha_1,\alpha_2,\beta_1,\beta_2)}(\Psi)$.
Introducing the variable
$t=x\, n^\beta$ in the integral, and using the fact that $\Phi(t)$ has support in $[-1,1]$, we
obtain $$
(\delta \otimes \delta \otimes \delta \otimes \delta)_{(\alpha,\beta)}(\Psi) = 
\lim_{n\rightarrow \infty} \int_{-\infty}^{\infty} f_n^{(\alpha,\beta)}(t) \, dt,
$$
where
$$
f_n^{(\alpha,\beta)}(t) \equiv
\frac{1}{\pi^2n^{2\alpha-5\beta}}\frac{(\Phi(t))^2\Psi(t/n^\beta)}{(t^2+1/n^{2(\alpha-\beta)})^2}.
$$
At this point we use the LDCT. In fact, for any $\alpha$ and $\beta$ with $2\alpha\geq 5\beta$ we
find that $|f_n^{(\alpha,\beta)}(t)|\leq g(t)$, where $g(t)\equiv \frac{LM}{\pi^2F}|t|^{2(m-2)}$.
Here we have used the same notation introduced in \cite{bag1}, and we have called 
$M\equiv \sup_{t\in
]-1,1[}\exp\{\frac{1} {t^2-1}\}$ and $L\equiv \sup_{t\in ]-1,1[}|\Psi(t)|$. Of
course, $g(t)$ is integrable in $[-1,1]$ whenever $m$ assumes values bigger or equal to 2.
Moreover, the function $f_n^{(\alpha,\beta)}(t)$ converges pointwise, whenever $2\alpha\geq
5\beta$, to a function $f^{(\alpha,\beta)}(t)$ which is equal to zero if $2\alpha> 5\beta$ and to
$\frac{(\Phi(t))^2\Psi(0)}{\pi^2 t^4}$ if $2\alpha= 5\beta$. In these conditions the LDCT can be
applied and we get 
\be
    (\delta \otimes \delta \otimes \delta \otimes \delta)_{(\alpha,\beta)}(\Psi)  = \left\{
          \begin{array}{ll}
          \frac{1}{\pi^2}A_{2,4} \delta(\Psi),    &       2\alpha=5\beta \\
           0,   &       2\alpha>5\beta, 
       \end{array}
        \right. 
\label{45}
\en
where, of course, $m\geq 2$.

The $( \otimes )_{(\alpha,\beta)}$ multiplication of five delta functions is again
computed very simply. We have
\beano
& &(\delta \otimes \delta \otimes \delta \otimes \delta \otimes
\delta)_{(\alpha_1,\alpha_2,\beta_1,\beta_2)} (\Psi) =  \\
& &= \lim_{n\rightarrow \infty} \int_{-\infty}^{\infty} \delta(x) \, \delta_n^{(\beta_1)}(x)\,
\delta_{red}(x,\frac{1}{n^{\alpha_1}}) \delta_n^{(\beta_2)}(x)\,
\delta_{red}(x,\frac{1}{n^{\alpha_2}}) \, \Psi (x) \, dx=0,
\enano
using again the fact that $\Phi(0)=0$ whenever $m>0$.

We are now ready to generalize these results: let $l$ be a natural number. Therefore, for any
$\Psi \in \D(\R)$, 
\be
(\underbrace{\delta \otimes ...\otimes \delta}_{2l+1})_{(\alpha_1,..,\alpha_l,\beta_1,..,\beta_l)}
(\Psi) =0 
\label{46}
\en
for any choice of $\alpha_i$ and $\beta_i$ and for $\Phi$ given by (\ref{23bis}) with $m>0$. On
the other hand the multiplication of an even number, $2l$, of delta functions may give a non zero
(and finite!) result. It depends, in general, on $\alpha_i$ and $\beta_i$ with $i=1,2,..,l$, see
equations (\ref{33}) and (\ref{35}). As we have already discussed for $l=2$, it is actually enough
to put $\alpha_1=\alpha_2=..=\alpha_l=:\alpha$ and $\beta_1=\beta_2=..=\beta_l=:\beta$. We obtain
\bea
    \!\!\!\!\!\!\!&&(\underbrace{\delta \otimes ... \otimes \delta}_{2l})_{(\alpha,\beta)}(\Psi) 
\equiv \nonumber \\
 && \equiv (\underbrace{\delta \otimes ... \otimes
\delta}_{2l})_{(\alpha,..,\alpha,\beta,..,\beta)}(\Psi)
           = \left\{
          \begin{array}{ll}
          \frac{1}{\pi^l}A_{l,2l} \delta(\Psi),    &       l\alpha=(3l-1)\beta \\
           0,   &       l\alpha>(3l-1)\beta. 
       \end{array}
        \right. 
\label{47}
\ena
Obviously, $A_{l,2l}<\infty$ only if $m\geq 2$.

\vspace{5mm}

We now move to the second definition of the multiplication we have introduced in the last
Section. We show that also this method gives non trivial results. 

We know already that the multiplication of two delta functions is certainly well
defined,  since it coincides with the multiplication obtained following the first
procedure. In other words, we have 
\be
    (\delta \otimes \delta)^A_{(\alpha,\beta)}(\Psi) = (\delta \otimes
\delta)_{(\alpha,\beta)}(\Psi)   = \left\{
          \begin{array}{ll}
          \frac{1}{\pi}A_{1,2} \delta(\Psi),    &       \alpha=2\beta \\
           0,   &       \alpha>2\beta, 
       \end{array}
        \right. 
\label{48}
\en
and $m$ must be bigger or equal to 2.

When we consider three delta functions we obtain, from (\ref{39}),
\bea
\hspace{-13mm}&&( \delta \otimes \delta \otimes \delta )^A_{(\alpha,\beta)}(\Psi)= \nonumber \\
\hspace{-13mm}&&=\frac{1}{2} \lim_{n \rightarrow
\infty}  \int_{-\infty}^{\infty} \left[ b_1 (\delta_{n}^{(\beta)}(x))^2
\delta_{red}(x,\frac{1}{n^\alpha}) \!+ \!b_2 \delta_{n}^{(\beta)}(x)
(\delta_{red}(x,\frac{1}{n^\alpha}))^2 \right] \Psi(x)\, dx. 
\label{49}
\ena
We will not give here all the details of this computation, which are very similar to those
discussed above. The steps are, more or less, the same: we change the variable in the integrals
putting $t =xn^\beta$, we restrict the integration range due to the compact support of $\Phi(t)$,
and then we use the LDCT which can be applied under certain conditions on $\alpha$, $\beta$ and
$m$. For example, the first contribution in (\ref{49}) converges to a finite quantity whenever
$\alpha \geq 3\beta$ and for $m\geq 1$. On the contrary, the second contribution is surely
convergent for $\alpha \geq 2\beta$ and for $m\geq 4$. Collecting these results we obtain that, for
all $m\geq 4$, then
\be
    (\delta \otimes \delta \otimes \delta)^A_{(\alpha,\beta)}(\Psi)  = \left\{
          \begin{array}{ll}
          \frac{b_1}{\pi}A_{2,2} \delta(\Psi),    &       \alpha=3\beta \\
           0,   &       \alpha>3\beta, 
       \end{array}
        \right. 
\label{410}
\en
which is, in general, different from the analogous result, $(\delta \otimes \delta
\otimes \delta)_{(\alpha,\beta)}(\Psi)=0$, obtained using the first method.

To multiply four delta functions we refer to equation (\ref{310}). For any $\Psi \in \D(\R)$ we
have 
\bea
\hspace{-8mm}& &(\delta \otimes \delta \otimes \delta \otimes \delta)^A_{(\alpha,\beta)}(\Psi)= \frac{1}{7}
\lim_{n \rightarrow \infty} \int_{-\infty}^{\infty} {\Large [} 2c_1 (\delta_{n}^{(\beta)}(x))^3
\delta_{red}(x,\frac{1}{n^\alpha}) + \nonumber \\ 
\hspace{-8mm}& &+ 3c_2 (\delta_{n}^{(\beta)}(x))^2
(\delta_{red}(x,\frac{1}{n^\alpha}))^2 + 2c_3 \delta_{n}^{(\beta)}(x)
(\delta_{red}(x,\frac{1}{n^\alpha}))^3  {\Large ]} \, \Psi(x) \, dx. 
\label{411}
\ena
Now we need to estimate, using the usual techniques, three different contributions: the first
 is convergent whenever $\alpha\geq 4\beta$ and for any natural $m$. The second one converges
whenever $2\alpha\geq 5\beta$ and $m\geq 2$. The last term, finally, converges if $\alpha\geq
2\beta$ and $m\geq 6$. We conclude that, for any $m\geq 6$, 
\be
    (\delta \otimes \delta \otimes \delta \otimes \delta)^A_{(\alpha,\beta)}(\Psi)  = \left\{
          \begin{array}{ll}
          \frac{2 c_1}{7\pi}A_{3,2} \delta(\Psi),    &       \alpha=4\beta \\
           0,   &       \alpha>4\beta. 
       \end{array}
        \right. 
\label{412}
\en
It may be worthwhile to notice that the condition on $m$ does not follow from the
requirement of $A_{3,2}$, to be finite. In fact $A_{3,2}<\infty$ for any natural
$m$. It follows from the analogous requirement for $A_{1,6}$, which appears in the computation of
the last contribution in (\ref{411}), the one proportional to $c_3$.

Of course an extra degree of freedom is present now: the coefficients $b_1$ in (\ref{410}) and
$c_1$ in (\ref{412}) must satisfy only the very weak constraints: $b_1+b_2=2$ and
$2c_1+3c_2+2c_3=7$. But, since $b_2, c_2$ and $c_3$ do not appear at all, any choice of $b_1$ and
$c_1$ is allowed.

We now generalize the above results. In general we get
\be
    (\underbrace{\delta \otimes ...\otimes \delta}_{l})^A_{(\alpha,\beta)}(\Psi)  = \left\{
          \begin{array}{ll}
          \frac{d}{\pi}A_{l-1,2} \delta(\Psi),    &       \alpha=l\beta \\
           0,   &       \alpha>l\beta, 
       \end{array}
        \right. 
\label{413}
\en
where $d$ is a positive constant and $m\geq 2(l-1)$. Again, this constraint on $m$ follows from a
term which, under these hypotheses on $m$, $\alpha$ and $\beta$ does not contribute to the final
result, that is, the one proportional to $\int_{-\infty}^{\infty}\delta_{n}^{(\beta)}(x))
(\delta_{red}(x,\frac{1}{n^\alpha}))^{l-1} \psi(x)\, dx$.

  \vspace{3mm}

\noindent{\bf Remarks }.--
(a) It is interesting to observe that, for any odd integer $n$, the $(\otimes)^A_{(\alpha,\beta)}$
multiplication of $n$ delta functions may be different from zero while, the analogous
computation made using  $(\otimes)_{(\alpha,\beta)}$ returns necessarily zero.

(b) It is straightforward to generalize all the results obtained in this Section even to the
multiplication of the derivatives of the delta function. The technique is, more or less, the same.
We refer to \cite{bag1} for the details on the regularization procedures of the distributions
$\delta^{(p)}(x)$.

\vspace{3mm}

As in \cite{bag1} we can apply these results to one dimensional physical models which describe
media with impurities localized in certain fixed points, or to the discussion of the classical
limit of a certain quantum mechanical situation. 
 Let us consider, for instance, a three-particles system described by a
factorazible wave function
$$
\Phi^{\epsilon}(x_1,x_2,x_3,t)=\Phi^{\epsilon}_1(x_1,t)\Phi^{\epsilon}_2(x_2,t)\Phi^{\epsilon}_3(x_3,t)
$$
where
$$
|\Phi^{\epsilon}_1(x,0)|^2=|\Phi^{\epsilon}_2(x,0)|^2 =|\Phi^{\epsilon}_3(x,0)|^2  \equiv
\frac{\exp\{-(x/\epsilon)^2\}}{\epsilon \sqrt{\pi}}.
$$
We know that $P_\epsilon (x_1,x_2,x_3) \equiv
|\Phi^{\epsilon}_1(x_1,0)|^2 |\Phi^{\epsilon}_2(x_2,0)|^2 |\Phi^{\epsilon}_1(x_3,0)|^2dx_1 \,
dx_2\, dx_3$ is the probability of finding at $t=0$ particle $i$ between $x_i$ and $x_i+dx_i$,
$i=1,2,3$, \cite{mer}. In the limit $\epsilon \rightarrow 0$ we get $|\Phi^{\epsilon}_i(x,0)|^2
\rightarrow \delta (x)$ (for instance in ${\cal D'}$), so that
 $P_\epsilon (x_1,x_2,x_3)
\rightarrow \delta (x_1)\, \delta (x_2)\, \delta (x_3)\, dx_1 \, dx_2\, dx_3$. Because of this we
say that $\epsilon \rightarrow 0$ corresponds to the classical limit of the
system: in fact each particle is sharply centered in a single point.

We may look, therefore, for the probability of finding the three particles
at the same point $x$, in this classical limit. Of course simple physical
considerations require this probability to be zero. Therefore, since this
probability should be proportional to $(\delta (x))^3$, we conclude that the
natural regularization is the one in (\ref{46}) with $l=1$ and for any choice of $\alpha$ and
$\beta$, or the one in (\ref{410}) with $\alpha >3\beta$ and $m\geq 4$.

\section{Another example: Klein-Gordon model in $1+1$ dimensions}

The model of free bosons which we are going to discuss in this Section is defined
by the following second order differential equation
\be
\left( \frac{\partial^2}{\partial t^2}-\frac{\partial^2}{\partial x^2}+m^2 \right)
\varphi (x,t)=0
\label{51}
\en
and by the equal time canonical commutation relations
\bea
\left\{
 \begin{array}{ll}
& [\varphi(x,t), \varphi(x',t)]=0 \\
& [\dot \varphi(x,t), \dot \varphi(x',t)]=0, \\
& [\varphi(x,t), \dot \varphi(x',t)]=i\delta (x-x').\\
  \end{array}
\right.
 \label{52}
\ena
Following the notation and the main steps of \cite{bd}, we expand the solution of the
Klein-Gordon equation in plane waves, 
\be
\varphi(x,t) =\int_{-\infty}^{\infty}\frac{dk}{\sqrt{4\pi \omega_k}} \left[ a(k)
e^{ikx-i\omega_kt}+ a^{\dagger}(k) e^{-ikx+i\omega_kt} \right],
\label{53}
\en
where $\omega_k =\sqrt{k^2+m^2}$ and the operators $a(k)$ and its hermitean conjugate 
$a^{\dagger}(k)$ are the coefficients of the expansion. 
They satisfy these canonical commutation relations:
\be
[a(k),a(k')]=[a^{\dagger}(k),a^{\dagger}(k')]=0, \quad [a(k),a^{\dagger}(k')]=
\delta(k-k').
\label{54}
\en

Let us call $\Psi_0$ the ground state of the theory, \cite{bd}. This is defined by
requiring that  $a(k) \Psi_0 =0 \: \: \forall k\in \R$. 

Interesting quantities to compute are the
expectation values in $\Psi_0$ of the field $\varphi(x,t)$ and of the product of the
field, $\varphi(x,t)\varphi(x',t')$. As in the four-dimensional situation, even
in this simpler model problems arise when we try to compute the matrix element of the product
$\varphi(x,t)\varphi(x,t)$. In particular we observe that $$
(\Psi_0, \varphi(x,t) \Psi_0)  =0,
$$
while, a straightforward calculation shows that
\be
\Delta_+(r_x-r_y)\equiv (\Psi_0, \varphi(r_x) \varphi(r_y) \Psi_0) =\int_{-\infty}^{\infty}
\frac{dk}{4\pi \omega_k} e^{-i\hat k \cdot(r_x-r_y)},
\label{55}
\en
where $r_x=(x_0,-x)$ and $\hat k \cdot(r_x-r_y)=\omega_k(x_0-y_0)-k(x-y)$. It is therefore evident
that, in the limit $r_x \rightarrow r_y$, $\Delta_+(r_x-r_y)$ diverges logaritmically. 
(Recall that in four dimensions the analogous divergence is
quadratic.)

Now we are ready to discuss the application of the regularizations proposed to the 
Klein-Gordon field. In particular we will discuss first the regularization of
$\varphi(x,t)$ when $t$ is considered an extra parameter. This choice is necessary, at
this stage of knowledge, since the analytical regularization has been introduced only in
$\R$, while the sequential completion method is formulated
in $\R^{ n}$. The generalization of the analytic regularization to $n>1$ is discussed in
\cite{breme2}. Even if it is easily seen that both the regularization procedures work
well as far as the smearing of the field is concerned, we will also conclude that the
multiplication discussed in Section 2 does not allow to control the divergence of
$\Delta_+(0)$, even if the time is considered properly and not as a parameter. We hope to
be able to reconsider positively this problem in a future paper. 

We start considering the analytic regularization of the field $\varphi$. Using definition
(\ref{22}) and considering $t$ as a parameter we get for any $\epsilon >0$,
\beano
& &\varphi^0(x+i\epsilon,t) \equiv \frac{1}{2\pi
i}\int_{-\infty}^{\infty}\frac{\varphi(y,t)\, dy}{y-(x+i\epsilon)} = \\
& &=\int_{-\infty}^{0}\frac{dk}{\sqrt{4\pi \omega_k}}a^\dagger(k)
e^{-ikx+i\omega_kt} e^{k\epsilon}+ \int_{0}^{\infty}\frac{dk}{\sqrt{4\pi
\omega_k}}a(k) e^{ikx-i\omega_kt} e^{-k\epsilon}, 
\enano
where some easy applications of the integration in the complex domain has been used.
Analogously we get
$$
\varphi^0(x-i\epsilon,t) =
- \int_{-\infty}^{0}\frac{dk}{\sqrt{4\pi \omega_k}}\, a(k)\, 
e^{ikx-i\omega_kt} e^{k\epsilon}- \int_{0}^{\infty}\frac{dk}{\sqrt{4\pi
\omega_k}}\, a^\dagger(k)\, e^{-ikx+i\omega_kt} e^{-k\epsilon}.
$$
Therefore the regularized function, $\varphi_{reg}(x,\epsilon;t)\equiv
\varphi^0(x+i\epsilon,t)-\varphi^0(x-i\epsilon,t)$, can be written as
\be
\varphi_{reg}(x,\epsilon;t) =\int_{-\infty}^{\infty}\frac{dk}{\sqrt{4\pi \omega_k}}
\left[a^\dagger(k)\, e^{-ikx+i\omega_kt}+ a(k)\,
e^{ikx-i\omega_kt}\right]\, P_{\epsilon}(k),
\label{56}
\en
where we have introduced the (even) function $P_{\epsilon}(k)\equiv
e^{-k\epsilon}\theta(k)+ e^{k\epsilon}\theta(-k)$.
From this equation and from (\ref{53}) it is easy to understand heuristically  why 
$\varphi_{reg}(x,\epsilon;t)$ is called a 'regularization' of $\varphi$: it appears evident,
in fact, that when $\epsilon \rightarrow 0$ then $\varphi_{reg}$  converges in some sense
to $\varphi$. This follows from the fact that, when $\epsilon \rightarrow 0$, hence
$P_{\epsilon}(k) \rightarrow \theta(k)+\theta(-k)$. Therefore, in this limit, this
function behaves like the unit function whenever considered 'inside an integral'. More precisely,
if $f(k)$ is an integrable function, then we have $$
\int_{-\infty}^{\infty}dk \, f(k) \lim_{\epsilon \rightarrow 0} P_{\epsilon}(k) =
\int_{-\infty}^{\infty}dk \, f(k).
$$

\vspace{5mm}

Let us now make this heuristical argument  rigorous, showing that $\varphi_{reg}(x,\epsilon;t)$
converges to $\varphi(x,t)$ in the topology of $\s'(\R)$ whenever $\epsilon$ is sent to 0.
In particular, we are going to show
that, in the limit $\epsilon \rightarrow 0$,  the following quantity 
\be
\delta_{\epsilon}(\varphi) \equiv  (\Psi_1,
\int_{-\infty}^{\infty}[\varphi_{red}(x,\epsilon;t)-\varphi(x,t)] \, \zeta(x)\, dx \,
\Psi_2 ), 
\label{56bis}
\en
goes to zero. Here $\Psi_1, \Psi_2$ are vectors of the Hilbert space, and $\zeta(x)$ is a
function in $\s(\R)$.
 Using equations (\ref{56}) and (\ref{53}), expliciting the form of
$P_\epsilon(k)$ and introducing the functions $a_{12}(k) \equiv (\Psi_1, a(k) \Psi_2)$,
 $a_{12}^{\dagger}(k) \equiv (\Psi_1, a^{\dagger}(k) \Psi_2)$, and the Fourier
transform of $\zeta(x)$,  $$\tilde \zeta(k) =
\frac{1}{2\pi}\int_{-\infty}^{\infty} \zeta(x) e^{ikx}\, dx,$$
 we can write
$\delta_{\epsilon}(\varphi)$ as the sum of four contributions all, more or less, of
the same kind. The first contribution, for instance,  is proportional to
$$
\int_{0}^{\infty}\frac{dk}{\sqrt{2\omega_k}}(e^{-k\epsilon}-1)a_{12}(k)
e^{-i\omega_k t} \tilde \zeta(k).
$$
Since $\tilde \zeta(k)$ is a function of $\s$ and $a_{12}(k)$ is surely well behaved, we
can use LDCT  to conclude that
the above integral converges to zero when $\epsilon$ goes to $0$. We arrive to 
similar conclusions also for the
other three contributions in $\delta_{\epsilon}(\varphi)$. This implies that
$\varphi_{red}(x,\epsilon;t)$ converges to $\varphi(x,t)$ in $\s'$.

\vspace{4mm}

We discuss now the way in which a delta family can be used in the regularization of the scalar
field. As for the analytic method we consider the time as a parameter. Therefore we
have \be
\varphi_n^{(\beta)}(x,t) \equiv \int_{-\infty}^{\infty} \delta_n^{(\beta)}(y)
\varphi(x-y,t) \, dy= \int_{-\infty}^{\infty}\Phi(q) \varphi(x-\frac{q}{n^\beta},t)\, dq,
\label{57}
\en
where, as usual, we indicate with $\Phi(x)$  the function generating the
$\delta$-sequence. We now prove explicitly that if $\Phi$ satisfies the following three
conditions, then $\varphi_n^{(\beta)}(x,t) \rightarrow \varphi(x,t)$ in $\s'$:

i) $\int_{-\infty}^{\infty} \Phi(x)\, dx =1;$

ii) $\Phi(x) =\Phi(-x);$ 

iii) $\Phi \in \s(\R).$

Incidentally, we observe that such a
function generates a delta family of Dirichelet type by means of the
procedure discussed in Section 2. Condition ii), which is not required in the original
definition of the functions of this family, is only an useful technical requirement. 

Since $t$ is considered as an extra parameter, we need to prove explicitly the
convergence of $\varphi_n^{(\beta)}(x,t)$ to $\varphi(x,t)$. For this reason, similarly to
what we have done in (\ref{56bis}), we  compute the following limit $$
\lim_{n\rightarrow \infty} \tilde \delta_{n}(\varphi) \equiv  \lim_{n\rightarrow \infty}
(\Psi_1, \int_{-\infty}^{\infty}[\varphi_n^{(\beta)}(x,t)-\varphi(x,t)] \, \zeta(x)\, dx
\, \Psi_2 ), 
$$
where $\Psi_1, \Psi_2$ and $\zeta(x)$ are the same as in $\delta_\epsilon(\varphi)$.
Using the parity of the function $\Phi$ and introducing again the functions $a_{12}(k)$ and
$a_{12}^{\dagger}(k)$, we can write $\tilde \delta_{n}(\varphi)$ as the sum of two
 contributions with the same structure. In particular the first term of 
$\tilde \delta_{n}(\varphi)$,  $\tilde \delta_{n,1}(\varphi)$, is 
$$
\tilde \delta_{n,1}(\varphi) = \int_{-\infty}^{\infty} \frac{dk}{\sqrt{4\pi \omega_k}}
\, a_{12}(k)\, e^{-i\omega_kt} \tilde \zeta(-k) \left( 2\pi
\tilde\Phi\left(\frac{k}{n^\beta}\right)-1\right).
$$
Again, we make use of the LDCT. The procedure is
now a bit tricky. First of all, since $\tilde \zeta$ belongs to $\s$, as well as
$\tilde \Phi$, it is clear that, $\forall n \in \N$, the function 
$$f_n(k) \equiv \frac{1}{\sqrt{4\pi \omega_k}}
\, a_{12}(k)\, e^{-i\omega_kt} \tilde \zeta(-k) \left( 2\pi
\tilde\Phi\left(\frac{k}{n^\beta}\right)-1\right)
$$
belongs to ${\cal L}^1(\R)$. In fact we can write $|f_n(k)|\leq g(k)$, with
$$
g(k)\equiv \frac{1}{\sqrt{4\pi \omega_k}}
\, |a_{12}(k)\,  \tilde \zeta(-k)| ( 2\pi M+1).
$$
Here $M$ is the supremum of the function $\tilde\Phi$. Obviously, since $g(k) \in {\cal
L}^1(\R)$, then also $f_n(k) \in {\cal L}^1(\R)$.

This implies
that, for any $\epsilon >0$, it is possible to choose a positive quantity $R_\epsilon$,
independent on $n$, such that $\int_{|k|>R_\epsilon}|f_n(k)|\, dk <\epsilon$. Due to the hypothesis
i) of normalization of the function $\Phi$, which can also be written in terms of its Fourier
transform as $2\pi \tilde \Phi(0) =1$, we deduce that, as far as $|k|\leq R_\epsilon$,
 $f_n(k)$ surely converges almost everywhere to the function zero. This means that,
using LDCT 
$$
\lim_{n\rightarrow \infty}\int_{|k|\leq R_\epsilon}|f_n(k)|\, dk = 
\int_{|k|\leq R_\epsilon}\lim_{n\rightarrow \infty}|f_n(k)|\, dk =0,
$$
which also implies that, given $\epsilon$, it exists $n_\epsilon \in \N$ such that, for
all $n>n_\epsilon$, $\left| \int_{|k|\leq R_\epsilon}|f_n(k)|\, dk \right| <\epsilon$.
We can conclude that for all $\epsilon >0$, it exists a natural $n_\epsilon$ such
that, for all $n$ bigger than $n_\epsilon$, 
$$
\left| \int_{-\infty}^{\infty}f_n(k)\, dk \right| <2\epsilon.
$$
An analogous estimate can be performed also for the second contribution of 
$\tilde \delta_{n}(\varphi)$,  $\tilde \delta_{n,2}(\varphi)$. We conclude  that
$\varphi_n^{(\beta)}$ converges to $\varphi$ in $\s'$.

Defining the following set of functions
\be
{\cal Z} \equiv \left\{ \Psi(x) \in \s(\R): \:\: \int_{-\infty}^{\infty} \Psi(x) \,
dx=1, \: \Psi(x)=\Psi(-x) \right\},
\label{58}
\en
we can summarize the above results in the:
\vspace{4mm}

\noindent
	{\bf Proposition 2}.
For all $\Phi \in {\cal Z}$ the function $\varphi_n^{(\beta)}=\int_{-\infty}^{\infty}
\Phi(q) \, \varphi(x-q/n^\beta,t)\, dq$ converges to $\varphi(x,t)$ in the topology of
$\s'$.

More results on this convergence will be discussed in the Appendix.

Once we have shown how the regularizations work for the quantum free field, we may think
to use $\varphi_n^{(\beta)}(x,t)$ and $\varphi_{red}(x,\epsilon;t)$ to
 eliminate (some) divergences
appearing in the quantum model. For instance we may think that the regularization of
$\Delta_+(r_x-r_y)$ can be made finite for $r_x=r_y$. Unfortunately this is not so. In fact, let
us define, as it is natural, \be
[\Delta_+(r_x-r_y)]_{(\alpha,\beta)} \equiv (\Psi_0, (\varphi(r_x) \otimes
\varphi(r_y))_{(\alpha,\beta)}  \Psi_0), 
\label{59}
\en
and let us focuse our attention in particular to $[\Delta_+(0)]_{(\alpha,\beta)}$.

We start computing, see (\ref{33}),
$$
I_n(\varphi)\equiv  (\Psi_0, \varphi_n^{(\beta)}(x,t)
\,\varphi_{reg}(x,\frac{1}{n^\alpha};t) \Psi_0),
$$
and then we discuss the limit of $I_n(\varphi)$ for $n$ diverging.
Considering only the non vanishing contributions ($a(k) \Psi_0= \Psi_0 a^\dagger(k) =0$),
and using the commutation relations of the bosonic operators $a(k)$ and 
$a^\dagger(k)$, we get
\be
I_n(\varphi) = \frac{1}{4\pi} \int_{-\infty}^{\infty} dw \, \Phi(w)
\int_{-\infty}^{\infty}\frac{dk}{\omega_k}e^{-ikw/n^\beta}\, e^{-\theta(k)k/n^\alpha},
\label{510}
\en
where $\theta(k)$ is a function which is equal to $1$ for $k\geq 0$ and to $-1$
otherwise.

Using the fact that, since $\Phi(x)$ is taken in ${\cal Z}$ then $\Phi(x)$ is 
 an even function, as well as its Fourier transform, we have 
\be
I_n(\varphi) = \frac{1}{2} \int_{-\infty}^{\infty}
\frac{dk}{\omega_k}\tilde\Phi\left(\frac{k}{n^\beta}\right)\, e^{-\theta(k)k/n^\alpha}=
\int_{0}^{\infty} \frac{dk}{\omega_k}\tilde\Phi\left(\frac{k}{n^\beta}\right)\,
e^{-k/n^\alpha}.  
\label{511}
\en
Let us introduce now this new set of functions:
\be
\tilde{\cal Z}_0 \equiv \left\{ \tilde\Psi(k) \in \D(\R): \:\: 
2\pi \tilde \Psi(0)=1, \: \tilde \Psi(k)=\tilde\Psi(-k) \right\}.
\label{512}
\en
It is obvious that the Fourier anti-transform, $FT^{-1}$, of any function in $\tilde{\cal Z}_0$
belongs to ${\cal Z}$, since $\D \subset \s$. It may be useful to take $\Phi$ such that 
$\tilde\Phi \in \tilde{\cal Z}_0$, since in this way $I_n(\varphi)$ can be computed easily
using numerical techniques. With the change of variable $q=k/n^\beta$, calling again $M$
the supremum of the function $\tilde \Phi(k)$, and assuming that the support of $\tilde
\Phi(k)$ is the interval $[-1,1]$, we deduce that
\be
I_n(\varphi) = \int_{0}^{1} \frac{dq}{\sqrt{q^2+\frac{m^2}{n^{2\beta}}}}\,\tilde\Phi(q)
\, e^{-qn^{\beta-\alpha}} \leq M\,  \int_{0}^{1}
\frac{dq}{\sqrt{q^2+\frac{m^2}{n^{2\beta}}}}\,
 e^{-qn^{\beta-\alpha}}.  
\label{513}
\en

In order to get analytic informations on the asymptotic behavior of $I_n(\varphi)$ we
begin with an easy estimate which shows that the above integral cannot be
convergent for $n\rightarrow \infty$ whenever $\alpha \geq \beta$. This follows from the
following analytic estimate: since for $q\in [0,1]$ $e^{-qn^{\beta-\alpha}}\geq 
e^{-n^{\beta-\alpha}}$, it follows that   \beano
\int_{0}^{1}\frac{dq}{\sqrt{q^2+\frac{m^2}{n^{2\beta}}}}\,e^{-qn^{\beta-\alpha}}
\!\!\!\!&&\!\!\!\!\geq  e^{-n^{\beta-\alpha}} \,
\int_{0}^{1}\frac{dq}{\sqrt{q^2+\frac{m^2}{n^{2\beta}}}} = \\
&&=e^{-n^{\beta-\alpha}}\log\left(\frac{n^\beta+\sqrt{n^{2\beta}+m^2}}{m}\right). 
\enano
Of course, whenever $\alpha \geq \beta$ the right hand side diverges. This does not
really imply that also $I_n(\varphi)$ diverges, as it is clear. Nevertheless it is a
very strong indication which, moreover, it is also supplemented by the following
remark: when $n\rightarrow \infty$, the first integral in (\ref{513}) behaves, when 
$\alpha \geq \beta$, like $\int_0^1\frac{dq}{q}\, \tilde\Phi(q)$, which can be finite only if 
$\tilde\Phi(q)$ goes to zero when $q\rightarrow 0$. This is not what we have since
$\Phi$ belongs to ${\cal Z}$, so that its value in $k=0$ is
$\frac{1}{2\pi}$. These result suggests that for $I_n(\varphi)$ to be converging, $\beta$
must be chosen bigger than $\alpha$. But also in this case it is not easy  to find an
analytic estimate for the integral in (\ref{513}) proving that
$\lim_{n\rightarrow \infty}I_n(\varphi)<\infty$. For this reason we have used
numerical procedures to compute this integral, for different choices of $\alpha$ and
$\beta$. Unluckily these numerical results seem to show again that  $\lim_{n\rightarrow
\infty}I_n(\varphi)=\infty$, even if the divergence is very slow.

\vspace{3mm}

Before ending this Section, we briefly comment on the complete regularization of the field, that is
the one in which we consider properly $t$ as the time coordinate of the field. First of all
we notice that, in a certain sense, only in this case we are allowed to speak of a {\em
canonical regularitazion} of the field since the general theory says that the
two-dimensional convolution $\varphi*\delta_n^{(\beta)}$ is a $C^\infty$ function and that
the two-dimensional analytic regularization is an analytic function. The computation of
$I_n(\varphi)$ does not present many differences with respect to the situation discussed
above and, by the way, the conclusion is still the same: we get, for any choice of
$\alpha$ and $\beta$, $\lim_{n\rightarrow\infty} I_n(\varphi)=\infty$. For this reason we
believe it is not worthwhile to give here the details of this procedure,
which are much heavier than those discussed above and, again, do not lead to a positive
conclusion.

\section{Conclusions}

In this paper we have discussed different generalizations of the multiplication of distributions
first introduced in \cite{bag1}. In particular, we have proposed possible modifications of the
sequential completion method which may be of some utility depending on the distribution to be
regularized. 

Furthermore, we have introduced two different definitions of multiplications of $N>2$
distributions, both of which generalize the definition given in Section 2 for $N=2$. Of course,
many other generalizations are also possible. We have shown how both these definitions can be used
to define the multiplications of an arbitrary number of delta functions localized all in the same
point. A quantum mechanical physical example has been also sketched.

Finally we have discussed a naive possibility of using our strategy in QFT. We have
shown that it is possible to regularize the quantum field in many ways, but unfortunately
we have also shown that the definition proposed in Section 2 does not allow to cancel out
the divergence appearing already for a free theory. Our future project are therefore to
look for some refinement of the procedure which allows to overcome this last problem. If
this new technique can be found, we can also try to extend the theory  to four-dimensional
models and to discuss the 
divergences coming from the Feynman graphs. The final aim is to consider a non abelian
gauge theory like QCD, \cite{muta}.

In this analysis we expect that a crucial role will be played by the function $\Phi$
and by the parameters $\alpha$ and $\beta$ which fix the multiplication. They should have the same
role as the free parameters in renormalization theory, whose values are fixed by the experimental
data.

\vspace{8mm}

 \appendix

\renewcommand{\theequation}{\Alph{section}.\arabic{equation}}

 \section{\hspace{-.7cm}ppendix :  A Convergence Remark}

In this Appendix we prove in a different (and easier) way that, whenever $\Phi(x) \in {\cal Z}$,
then  $\varphi_n^{(\beta)}(x,t) \rightarrow \varphi(x,t)$ in $\s'$.

Let $\zeta(x) \in \s(\R)$. After some easy computation we deduce that
\beano
A_n^\beta &\equiv &\int_{-\infty}^\infty \varphi_n^{(\beta)}(x,t)\,\zeta(x)\, dx\, -
\int_{-\infty}^\infty \varphi(x,t)\,\zeta(x)\, dx= \\
&&= \int_{-\infty}^\infty\left(\delta_n^{(\beta)}(y)-\delta(y)\right)\, \int_{-\infty}^\infty
\varphi(x-y,t)\, \zeta(x)\,dx\,dy.
\enano
Of course the integral $\eta(y) \equiv\int_{-\infty}^\infty
\varphi(x-y,t)\, \zeta(x)\,dx$ is continuous in $y$. Using the results in \cite{kor},
 we can conclude that $A_n^\beta\rightarrow 0$ for $n\rightarrow \infty$. In fact, in
particular, if $\Phi(x)$ is taken positive, then it generates a delta family of positive type, so
that for any function $f(x)$ continuous in the origin we have  $$
\int_{-\infty}^\infty \delta_n^{(\beta)}(x) f(x) \,dx \rightarrow f(0).
$$
If $\Phi$ is not positive the same conclusion still holds since $\eta(y)$ is also differentiable
in $y=0$, see \cite{kor}.

\vspace{50pt}

\noindent{\large \bf Acknowledgments} \vspace{5mm}

	It is a pleasure to thank Dr. F. Guttuso for his precious help in finding
some bibliographic material. Thanks are also due to Dr. R. Belledonne for her kind
reading of the manuscript. We also thank M.U.R.S.T. for its financial support.

\end{document}